# Atomic-scale Imaging of Iodide-Gold Interactions in Nanoconfined Liquid-Solid Interfaces

Oliver R. Waszkiewicz[a], Yuxiang Zhou[a], Baptiste Gault [a,b,c], Finn Giuliani[a], Mary P. Ryan[*a], Ayman A. El-Zoka[*a, c]

[a] Department of Materials, Royal School of Mines, Imperial College London, Exhibition Road, SW7 2AZ, UK

[b] Max Planck Institute for Sustainable Materials, Max-Planck-Str. 1, 40239 Düsseldorf, Germany

[c] present address Univ. Rouen Normandie, INSA Rouen Normandie, CNRS, Groupe de Physique des Matériaux, UMR 6634, F-76000, Rouen, France

[*]Corresponding Author

## Abstract

Functionalization of nanoporous metallic materials enables the tailoring of surface chemistry and morphology in nanostructured materials, optimising their performance for electrocatalytic and sensor applications. Liquid-phase chemical functionalization is governed by liquid-solid interfaces. Yet, these interfaces remain poorly understood due to the challenges of characterising the liquid phase at high spatial and chemical resolutions. To elucidate pathways for functionalizing nanoscale metals, it is crucial to measure the distribution of species, including light elements, across the liquid-solid interface, capturing both reactants and products. Here, we employ cryogenic atom probe tomography to directly analyse frozen liquid–solid reaction interfaces at near-atomic resolution. Focusing on the interaction of iodide and sodium ions with nanoporous gold, we observe the formation of iodine-containing complexes on gold nanoligament surfaces and sub-surfaces. These findings reveal aspects of the gold–iodide system that were previously hidden, including the reaction mechanism between iodide and gold atoms on the surface, and the multiple gold-iodide complexes forming. Our work demonstrates that cryogenic atom probe tomography can provide unprecedented visualisation and characterisation of nanoscale interfaces during chemical and electrochemical reactions, with potential implications for modern manufacturing, energy technologies, and sustainable materials development.

# 1. Introduction

Nanoporous metals (NPMs) made by controlled dealloying corrosion have shown the potential for accelerating the development of electrochemical systems, owing to the inherent high surface area to volume ratio, high electrical conductivity, tuneable pore size and surface chemistry[1–3], which serve to enhance reactivity for chemical, electrochemical, and photochemical applications [4,5]. For decades, research in this field has primarily focused on nanoporous gold (NPG), typically produced by dealloying silver–gold alloys. Since then, similar dealloying strategies have been extended to other metals, including copper [6] and platinum [5], motivated by their potential for specific electrocatalytic applications.

An important application of nanoporous metals (NPMs) is their use as nanostructured templates for creating novel surface chemistries and nanocomposites tailored to specific applications. NPG is particularly advantageous because gold is relatively more chemically stable than other metals. Electrodeposition of metals and metal oxides onto NPG has been demonstrated, including the deposition of metals such as copper [7,8], and metal oxides such as $MnO_2$ [9] within the pores and onto the nanoligament surfaces. A more straightforward route to functionalize NPG relies on the inherent chemical reactivity between gold and specific species, such as iodide, enabling direct surface modification.

Gold exhibits a strong chemical affinity for iodide ions, which have been reported to adsorb onto gold surfaces and form stable gold–iodine complexes [10–12]. This interaction modifies the surface chemistry and electronic properties of gold, influencing its reactivity and enabling targeted functionalization [13]. These properties make gold–iodide systems attractive for applications in catalysis, sensing, and electrochemical processes, where controlled surface modification is essential for optimising performance.

As in all (electro)chemical systems, the iodide-gold interface is the most influential component during the chemisorption and reaction of iodide with gold. At the location where the liquid meets solid, charge transfer between atoms at the electrodes' surfaces and interfacial species determines the reactions occurring along with their (electro)chemical pathways[14–16]. A fundamental understanding of how this reaction proceeds, serving to outline pathways for optimisation and application of gold-iodide and other systems, can only be achieved if our perception of liquid-solid interfaces can be advanced from indirect physiochemical signals to real-time quantitative analysis on a scale compatible with the reactive processes, i.e. at the nanometre or below.

Cyclic voltammetry, electrochemical impedance spectroscopy and spectro-electrochemical operando methods have provided critical yet indirect insights on reactive processes at liquid-solid interfaces, with signal originating from a macroscopic average over the whole electrode/interface[17–20]. Scanning probe microscopy has higher spatial resolutions and provides unique information on surface processes[21,22], but studying more morphologically complex or buried interfaces remains highly challenging. Large Keggin ions (radii ~ 1 nm) were imaged in the electric double-layer on gold nanorod surface by cryogenic transmission electron microscopy (cryo-TEM)[23]. Microanalytical capabilities of cryo-TEM remain limited, and the electron scattering by the liquid tends to limit the image resolution[24].

There is an analytical gap for a technique with the chemical sensitivity needed to elucidate electrochemical reaction mechanisms at the nano- or sub-nanoscale, leaving numerous open questions about reactive liquid-solid interfaces. Cryogenic atom probe tomography (cryo-APT) appears as a promising technique that may directly resolve atomic-scale electrochemical processes in 3D, including for light elements, ions, and molecules[25,26]. APT relies on the field evaporation of individual atomic or molecular ions from submicron needle-shaped specimens. Recent advancements in ultra-high-vacuum (UHV) cryogenic sample preparation [9,27–29] and transfer methods [30–32] have enabled APT studies of frozen liquids [33–35], liquid-solid interfaces [33,36,37], and other soft matter samples preserved in their native state [38–41]. Compared to the decades of research and development in cryo-EM [42–47], cryo-APT of frozen liquids is in its infancy.

El-Zoka et al. demonstrated how the fixation of heavy water was facilitated using nanoporous gold (NPG)[33] enabled the cryo-APT analysis of larger volumes of liquids, compared to earlier studies[48,49]. The use of NPG was further explored [50,51] and expanded into nanoporous copper (NPC). Challenges remain in interpreting cryo-APT, as frozen anions and cations in solution can be expected to behave very differently[33], leading to potential losses in detection or a complete lack of detection, resulting in a skewed representation of the system's ionic composition[33,52]. If routine and robust analysis of electrochemical liquid-solid interfaces is to be achieved in cryo-APT, quantitative detection of all frozen ionic species is required.

In this study, we expand our understanding of the functionalization of NPG with iodide using interfacial characterisation, aiming to investigate the possible extent of iodide anion and iodine-containing complex detection in cryo-APT. The successful application of cryo-APT here in characterising NPG-I shows prospects for investigations into other, more complex electrochemical systems that will advance fundamental electrochemistry.

## 2. Results & Discussion
## 2.1. Overview of the APT data

To fabricate the APT specimen, the "satellite dish" focused-ion-beam (FIB) milling methodology demonstrated by El-Zoka et al. [33] was used, shown in **Figure 1**. The NPG sample (pore size approx. 15 nm) carrying the aqueous solution was plunge-frozen in liquid nitrogen and milled until the ice-supported above the NPG was ~3 μm in height, and the nanoporous gold was less than 100 nm in radius. **Figure 1b** and **1c** show the geometry of the final specimen and the clearance required to perform APT analysis, while the close-up in **Figure 1d** shows the network of nanopores arising from dealloying extending up to 5 μm below the surface. This method creates an APT tip directly onto the substrate without requiring a cryogenic lift-out workflow [50], simplifying the process and reducing preparation time have enabled the detection of over 60 million ions in each specimen, which is now routine and allows statistically significant compositional analyses.

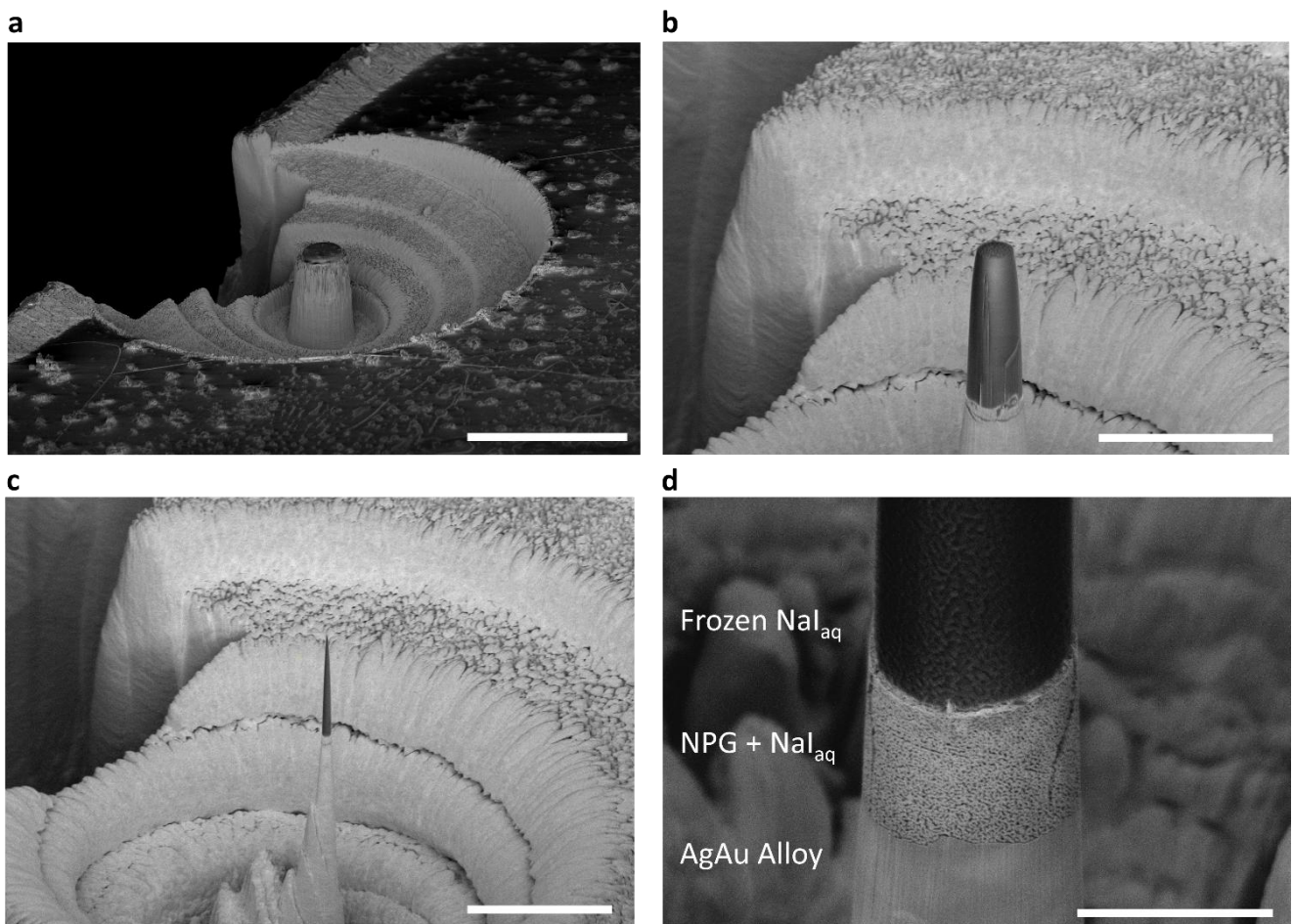

**Figure 1**: SEM Images of the cryo-APT sample preparation process of NPG and frozen NaI_{aq.} **a)** Showing the "satellite dish" milling pattern, with the concentric FIB milling patterns getting progressively smaller, and the clearance for the laser and local electrode in APT analysis (scale bar = 100 μm). **b,c)** The sample was gradually sharpened until the radius of the tip was less than 100 nm, and the height of the ice on top was ~3 μm. (scale bars 25 μm). **d)** Close to the final APT specimen showing the NPG morphology with the frozen NaI_{aq} inside the pores (scale bar = 5 μm)

The 3D atom maps of the major detected ions are shown in **Figure 2a**. The mass spectrum collected from within the 15nm NPG substrate (to the left of the red dotted line in **2a**) filled with the frozen 50 mM NaI_{aq} electrolyte is plotted in **Figure 2b**. A separate mass spectrum and reconstruction were created for the "bulk-electrolyte" region in **Figure S3**. Significant detection of Au and I complexes in the form of $AuI^{+/2+}$, $AuI_2^{+/2+}$, $Au_2I^{+/2+}$, and $AuI(H_2O)^{+/2+}$, which haven't been seen in previous studies of Au and Chloride solutions. Comparatively, no metallic complexes are present for Na, which is observed in the spectrum as $Na^+$ and $Na(H_2O)^{x+}$ up to a value of x = 4, which remains consistent with the previous studies, where an aqueous 50 mM NaCl electrolyte was used on the same substrate[33]. We detect both $I^+$ and $I^{2+}$ in the spectrum, stemming from the lower electron detachment energy of iodide ions compared to chloride ions[53]. Overall, this leads to remarkably higher and consistent detection of anions in frozen solutions in cryo-APT [33,52].  A composition profile across the liquid-solid interface is plotted in **Figure 2c,** showing an increase in the detection of solute ions and Au-I ions. It shows increased anion detection inside the nanopores (compared to in the bulk phase of the electrolyte on top of the NPG), with peaks corresponding to $I^+$, $I^{2+}$, $I(H_2O)^{+\cdot}$ and $I_2^+$.  The profile for $H_{2n+1}O_n^+$, which are water molecule clusters, was calculated by the addition of the respective ionic concentrations for each cluster.

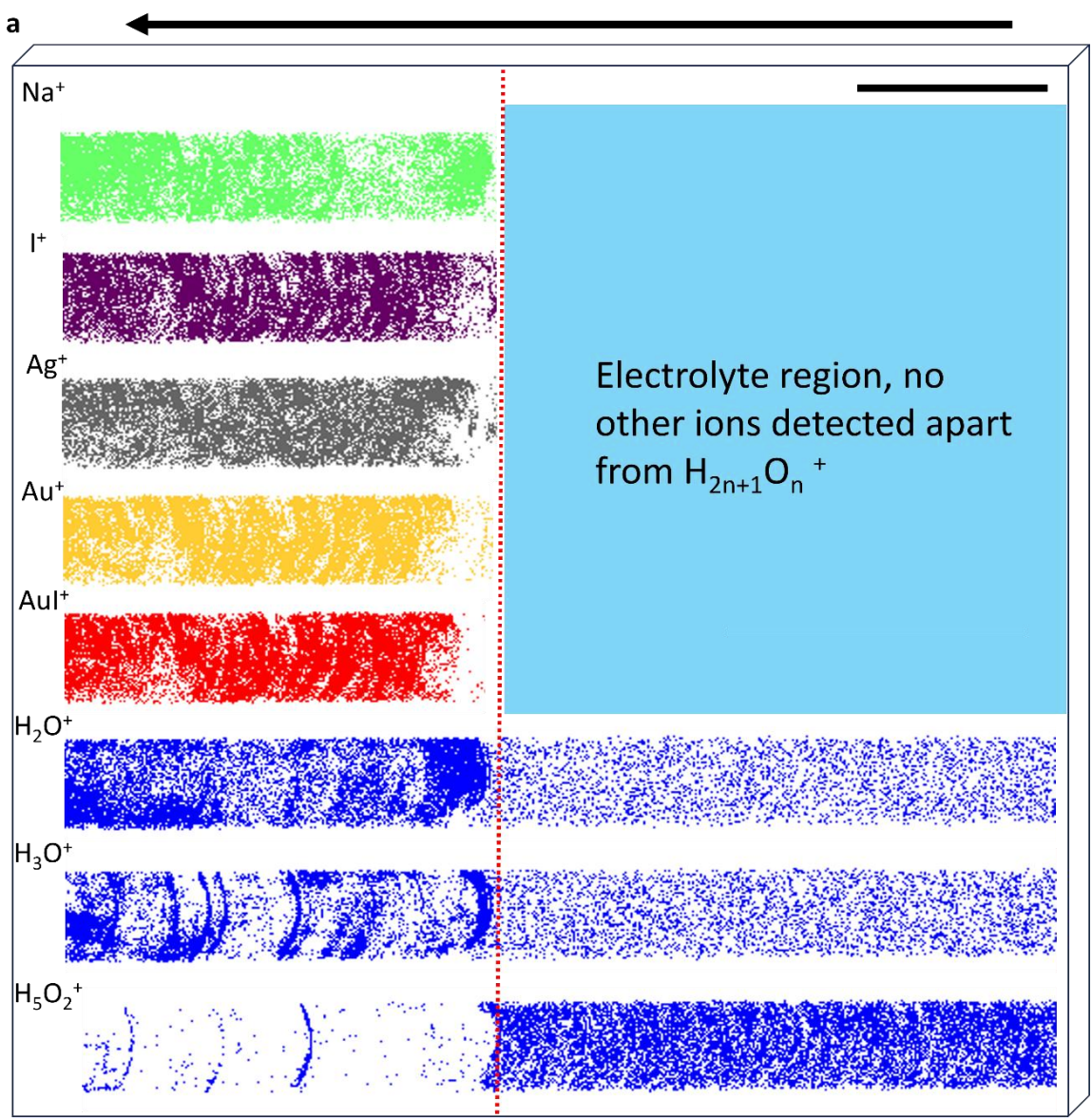

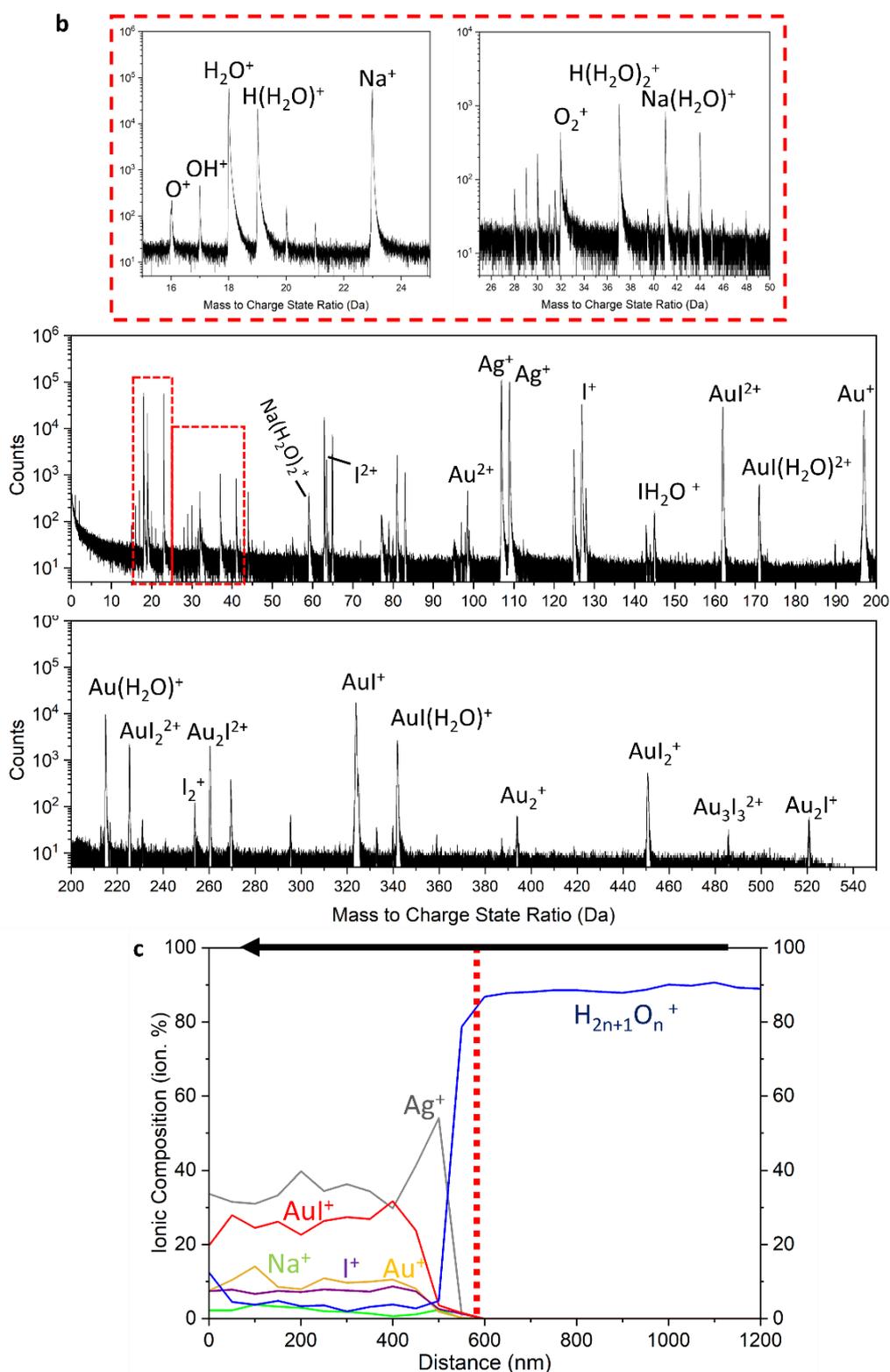

**Figure 2**: **a)** Atom maps of the main detected ions over the bulk liquid–NPG interface. The figure was reconstructed using a primary evaporation field of 24 V/nm (Ag) and an initial rip radius of 40 nm (Ag). **b)** Mass spectrum showing mass-to-charge state ratio from 0-550 Da collected from within the NPG sample, without thermal annealing, sample (left of the red dotted line) on a logarithmic plot. The major and of interest peaks have been labelled with their respective ions. The supplementary section contains the mass spectra collected for the annealed NPG samples (**Figure S4 and S5**). **c)** Global 1-dimensional concentration profile through the frozen $NaI_{aq}$ into the NPG region filled with $NaI_{aq}$, with the major interface depicted by the red dotted line. The direction of the profile for the atom maps is shown by the arrows above.

## 2.2. Detection of Solute Na and I in Pores

Reconstructions with increased magnification based on the mass spectrum from **Figure 2b(**to the left of the red dotted line) are shown in **Figure 3.** We can see the nanoporous morphology in the reconstruction, with the gold displayed as a solid, isosurface concentration of 10 ion% Au$^+$. These reconstructions, showing different projections, help us gain qualitative insight into the distribution of the detected species, while highlighting the ability to preserve the nanoporous morphology in the reconstructions despite the complex phenomena associated with field evaporation in systems with multiple phases/species.

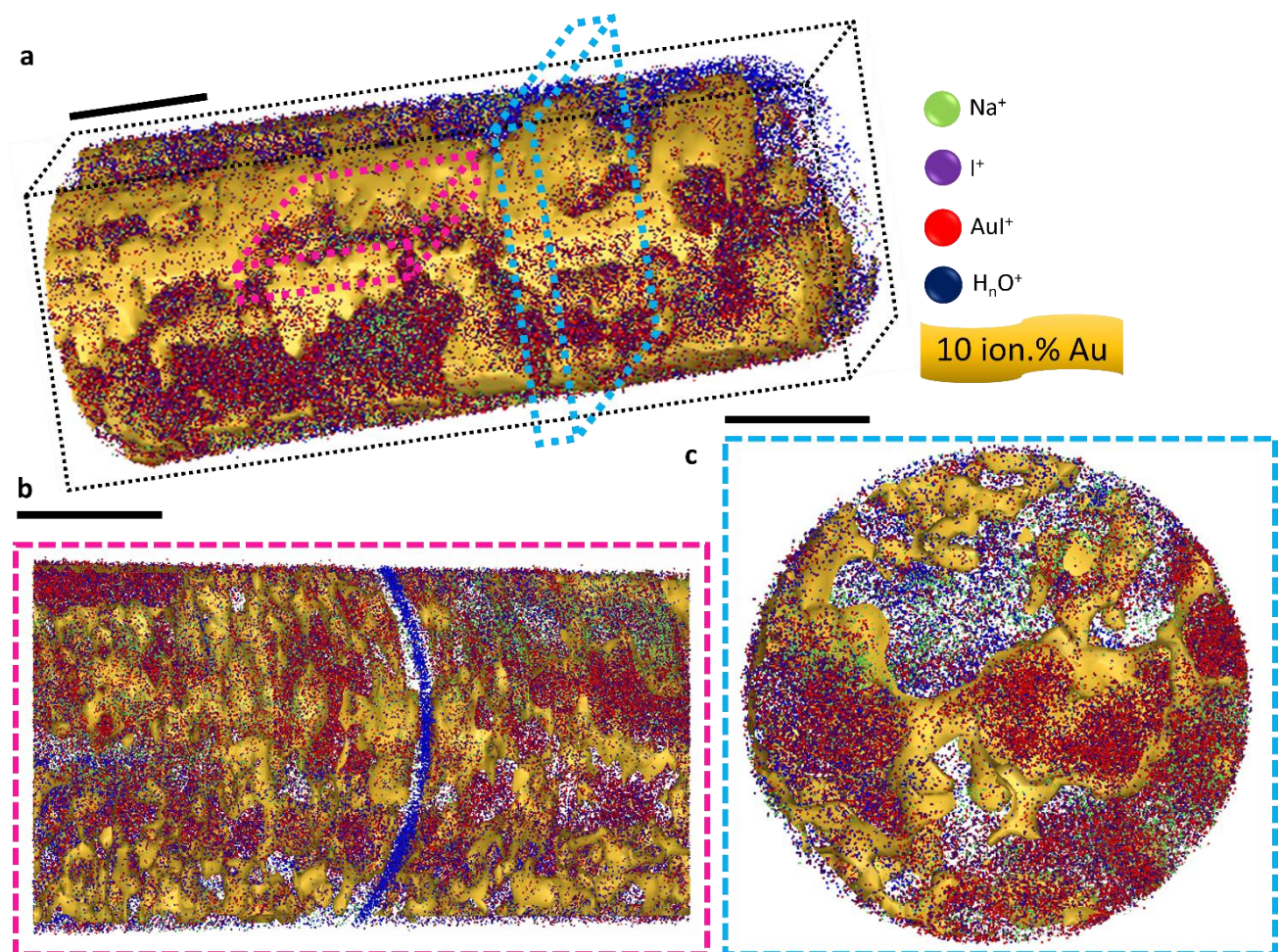

**Figure 3: a)** 3D reconstruction of the 15 nm NPG sample showing the primary ions of interest and a scale bar = 20 nm. **b)** Projection 10 nm thick slice through the tomogram in **a)** along the plane displayed (scale bar = 20 nm). **c)** Projection 10 nm thick slice through the tomogram in **a)** along the plane displayed. (scale bar = 20 nm). The figures were reconstructed based on a primary evaporation field of 24 V/nm (that of Ag) and an initial tip radius of 40 nm.

El-Zoka et al . [33] had already reported the detection of sodium ions only inside the nanoporous layer in an NPG-NaCl$_{aq}$ system. This observation was attributed to solute concentrating within the nanopores, driven by the anticipated capillary freezing-point depression effect. The solutions inside nanopores are expected to freeze at temperatures well below the freezing point and at a time later than the point of freezing of the solution on top of the nanoporous layer [54,55]. Another important aspect that might contribute to this increase in solute concentrations is the higher background signal in the pure frozen solution part of the specimen, ~200-500 ppm/ns, compared to the NPG region of the specimen, ~5ppm/ns, due to lower conductivity in ice, which might lead to masking signals of solute ions.

We expect minimal chemical affinity of Na$^+$ ions on the gold surfaces because sodium is a hard cation with a small ionic radius and a strongly bound hydration shell. According to the hard and soft acids and bases theory (HSAB)[56,57], Na$^+$ is a hard acid, while gold behaves as a soft acid, particularly at undercoordinated sites on the NPG surface[56,57]. This mismatch results in weak or negligible specific adsorption of Na$^+$ onto the gold surface. Furthermore, Na$^+$ remains strongly hydrated in aqueous solution, and the energy needed to either partially or fully desolvate it to facilitate the interaction with the surface is typically not compensated by favourable interactions with gold. While it could be pushed onto the surface during crystallisation due to partitioning into the liquid phase, we do not observe a typical clustering of Na near any nanoligament surfaces.

In contrast, iodide ions exhibit a strong chemical and electrochemical affinity to gold surfaces [58,59]. Iodide (I$^-$) is a soft, polarizable anion that can specifically adsorb onto gold, forming stable interactions due to a good match in softness per HSAB theory. At open-circuit potential in a 50 mM NaI solution, the gold surface may carry a slight positive charge[60] but will likely be modified by adsorbed iodide (I$^-$) to exhibit a negative surface charge. This local environment could attract $Na^+$ electrostatically, but this is still governed by long-range Coulombic forces rather than chemical bonding. Therefore, even in the presence of a surface charge, $Na^+$ remains a diffuse, non-specifically interacting ion, with limited entry into the nanopores unless driven by other forces (e.g., an applied potential or freezing). As a result of these factors, the concentration of iodide inside the NPG region is higher than that of sodium **(Figure 2c),** even though it is generally more challenging to detect anions in APT.

## 2.3 Formation of Au-I Complexes

The detection of Au–I-containing molecular ions raises the question of the formation of Au-I complexes on nanoligament surfaces. During dealloying, the less noble metal (in this case Ag) in the precursor alloy is selectively dissolved while the more noble metal (Au) undergoes surface diffusion, leading to the formation of a nanoporous architecture with bi-continuous nanoligaments [1,2,61]. The surface diffusion mechanism operating during dealloying results in an Au-rich shell and an Ag-rich core [8,62].

A 1D composition profile across a single nanoligament after immersion in 50mM NaI is plotted in **Figure 4a**, along with the corresponding 3D reconstruction based on the cylindrical region of interest (ROI). In this case, molecular ions are decomposed into their atomic constituents, i.e., an $AuI^+$ ion is counted as both an Au and an I atom. The element-specific decomposition reveals a spatial co-localisation of Au and I, both with similar concentrations, with the nanoligament "shell" comprising 40 at. % Au and 40 at. % I. We also do not detect any significant $I^+$ inside the pore or near nanoligaments despite low background counts in these regions (<4 ppms/ns), suggesting further that the reaction between Au and I is chemical and that the Au in nanoligaments is fully reacted with iodide at the surface and within the nanoligaments. A similar co-localisation between Ag and I is consistently not observed.

In **Figure 4b**, an ionic profile calculated across the liquid-solid interface at another pore reveals that the atomic Au and I profiles in **4a** are comprised mostly of ionic AuI$^+$ with Au$^+$ and I$^+$ only making up 20 ion.% each. Most of the expected Au near the surface of the ligament has reacted, becoming iodised. In many instances, we observe the Au-rich shell thickness to be ~4 nm, which is significantly larger than other shell thicknesses reported for dealloyed AgAu (~ 2 nm)[8]. As the chemical reaction between Au and I progresses, the diffusion of Au atoms at the surface may expose additional Au atoms at the sub-surface, leading to the reaction of the Au layers beneath and a possible increase in the Au-I shell thickness, although this cannot be confirmed with only APT. The AuI formed on the NPG, in the presence of iodide-containing solutions, is also prone to possible dissolution, as evidenced in previous studies on the dissolution of AuI in iodide solutions[63,64].

To confirm whether the AuI probed in **Figure 3a** is within the frozen liquid or a solid phase, additional APT experiments were performed. Following annealing at 200 °C for 30 minutes to grow the pores to approximately 35 nm, we first immersed the NPG in 50 mM NaI for 2 hours, then submerged it in ultra-pure water overnight, and finally performed cryo-APT analysis of the reacted NPG filled with water. As shown in the profile in **Figure 4c**, up to 20 ion. % of AuI is still detected in the nanopore shell, confirming that the AuI shell probed in **Figure 4a** is at least partially in solid phase, and that the remainder was dissolved in NaI and had been washed out of the pores.

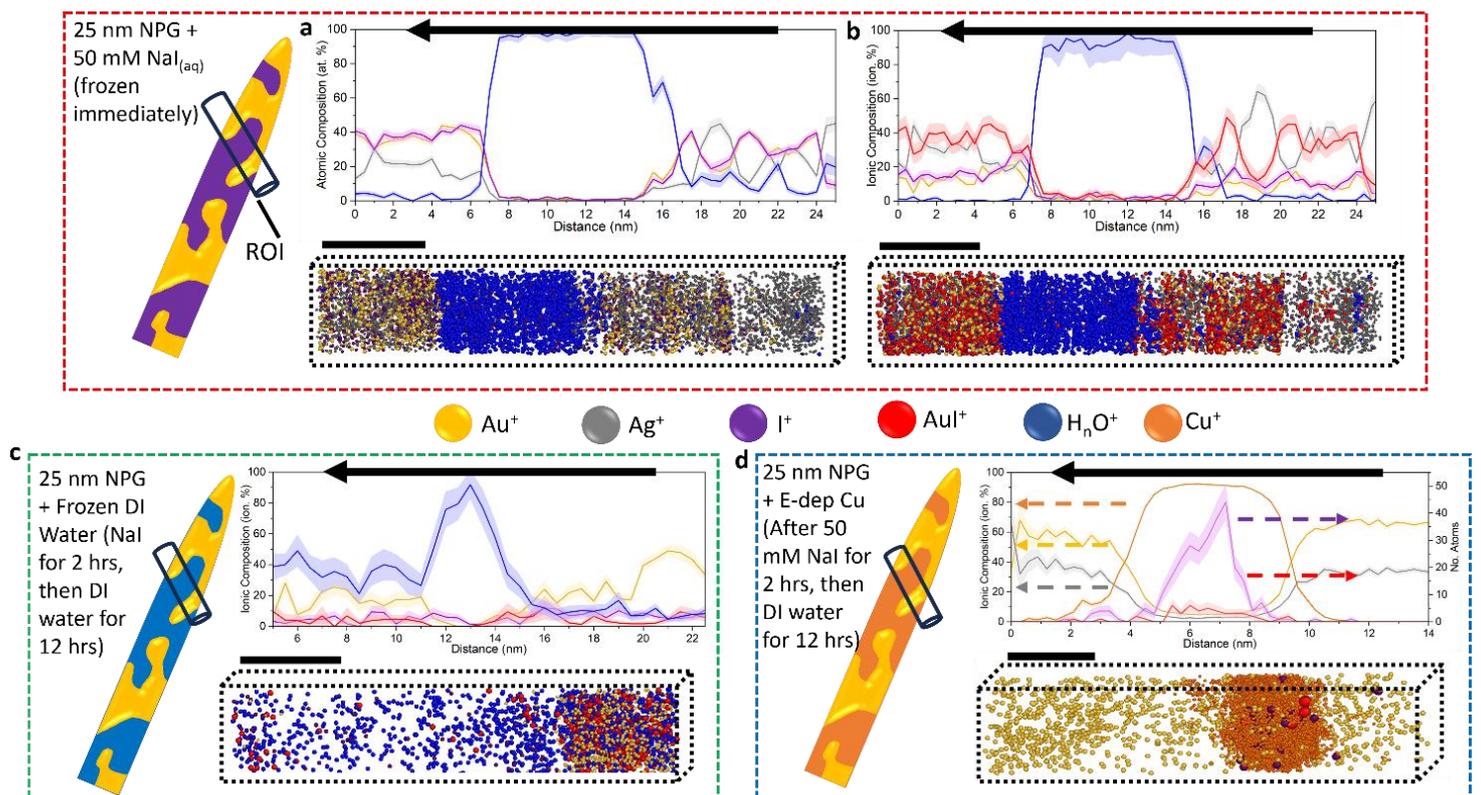

**Figure 4: a)** Concentration profile through a nanopore, and corresponding decomposed atomic 3D APT reconstruction showing the nanoligament chemistry of an NPG sample coarsened at 200 °C for 30 mins filled with 50 mM NaI$_{aq}$ (scale bar =5 nm). **b)** Concentration profile through a nanoligament surface from the same sample as **a)**, displaying the location of AuI over a ligament/pore interface, and the corresponding reconstruction (scale bar = 5 nm). **c)** Concentration profile through a nanopore, and corresponding 3D APT reconstruction showing the nanoligament chemistry of an NPG sample coarsened at 200 °C for 30 mins, submersed in 50 mM NaI$_{aq}$ for 2 hours, and then stored in DI water overnight (scale bar = 5 nm). **d)** Concentration profile through a nanopore, and corresponding 3D APT reconstruction showing the nanoligament chemistry of an NPG sample coarsened at 200 °C for 30 mins, submersed in 50 mM NaI$_{aq}$ for 2 hours, and then filled with electrodeposited copper, showing the location of AuI/I over a ligament/pore interface (scale bar = 3 nm). The schematics on the left of the subfigures show the history of the samples.

We then analyse NPG, after submersion in NaI$_{aq}$, following the established workflow with Cu deposition into the pores to form a solid composite[65]. The profile in **Figure 3d** shows that I$^+$ and AuI$^+$ are detected within the Cu filling rather than at the ligament surface, suggesting that the cathodic polarisation of the sample during deposition has caused the decomposition of Au and I. By freezing this Au-I reaction, cryo-APT could uniquely and directly resolve nanoscale reaction intermediates that are metastable and result from the interaction between Au and I at the ligament surface and in the subsurface region.

It is expected for the I species to sit at the surface of the ligament, which would be in line with strong chemisorption of I to Au in both gaseous and aqueous forms of the anion, seen through LEED and STM studies on Au surfaces[11,66]. In both previous studies, high iodide coverage on Au(111) was observed. Overall, our mass spectrum shows a 1:1 atomic ratio of Au and I, comprising Au, I, and AuI ionic species. In nanoporous metals, the higher the curvature of the ligaments [67] [ref], alludes to the higher density of low coordination surface sites on ligament surfaces, compared to the surface of the bulk metal [68,69]. Such low-coordination surface sites were shown to be the origin of NPG's catalytic activity in chemical and electrochemical reactions, due to higher binding energies even for weakly adsorbed species such as CO[70]. These well-documented trends suggest that the confinement of Au-rich pore surfaces and the enrichment of the ligament surface in low-coordination sites could play a role in the strong interaction between NPG and I, as directly characterised in this study. Furthermore, we observe passivation-type behaviour by measuring the open-circuit potential (OCP) over time as the NPG sample is immersed in the 50 mM NaI$_{aq}$ solution (**Figure S8**). This increase in the OCP indicates that the I- ions initially react with the gold ligands, forming a passivated AuI layer that reaches equilibrium after some dissolution, depending on the iodide concentration within the nanopores. As no current was allowed to flow across the cell during the measurements, the change in OCP will reflect chemical reactions/ changes in the interfacial structure rather than an electrochemically driven process.

## 2.4 Dissolution of Au-I Complexes

Our discussions in the previous subsection have shown that a considerable amount of the AuI detected is within the liquid phase. This suggests that the AuI formed at nanoligament surfaces is further dissolved in the remaining unreacted iodide inside the pores. In this section, we further investigate the dissolution behaviour. Proximity histograms (proxigrams) are average concentration profiles measured as a function of distance from an isosurface concentration[71]. **Figure 4** shows proxigrams and magnified Au and AuI species reconstructions across gold interfaces. They help quantify compositional change across interfacial regions with more complex geometries, reducing the bias that can arise from arbitrarily selecting regions for other 1D concentration profiles. In **Figure 5a**, NPG was immersed in a NaI electrolyte and frozen at cryogenic temperatures after 10 minutes. We again observe an AuI-rich region at the gold interface, extending slightly into a nanopore. The profiles for $AuI_2$ and $Au_2I$ closely correlate with that of AuI. It is also observed that the profiles of $AuI(H_2O)$ and $Au(H_2O)$ extend further into the porous region. While it's been observed in almost all metals, a hydrated species is seen on the surface during APT analysis, it is also possible that these hydrated Au species are dissolving from the AuI/Au surface [28,40,72]. Prolonged exposure to $NaI_{(aq)}$ (2 h immersion followed by DI rinsing overnight before freezing) results in a depletion of the AuI surface layer from 20 ion.% to 10 ion.% between **a** and **b**. At the same time, there is an increase in $Au(H_2O)^+$ at the surface of the ligament, with the ionic concentration profile remaining at an elevated level into the porous region. These trends are supported by the 3D reconstructions in **b** and **d**, revealing that extended electrolyte exposure drives the AuI species from the surface.

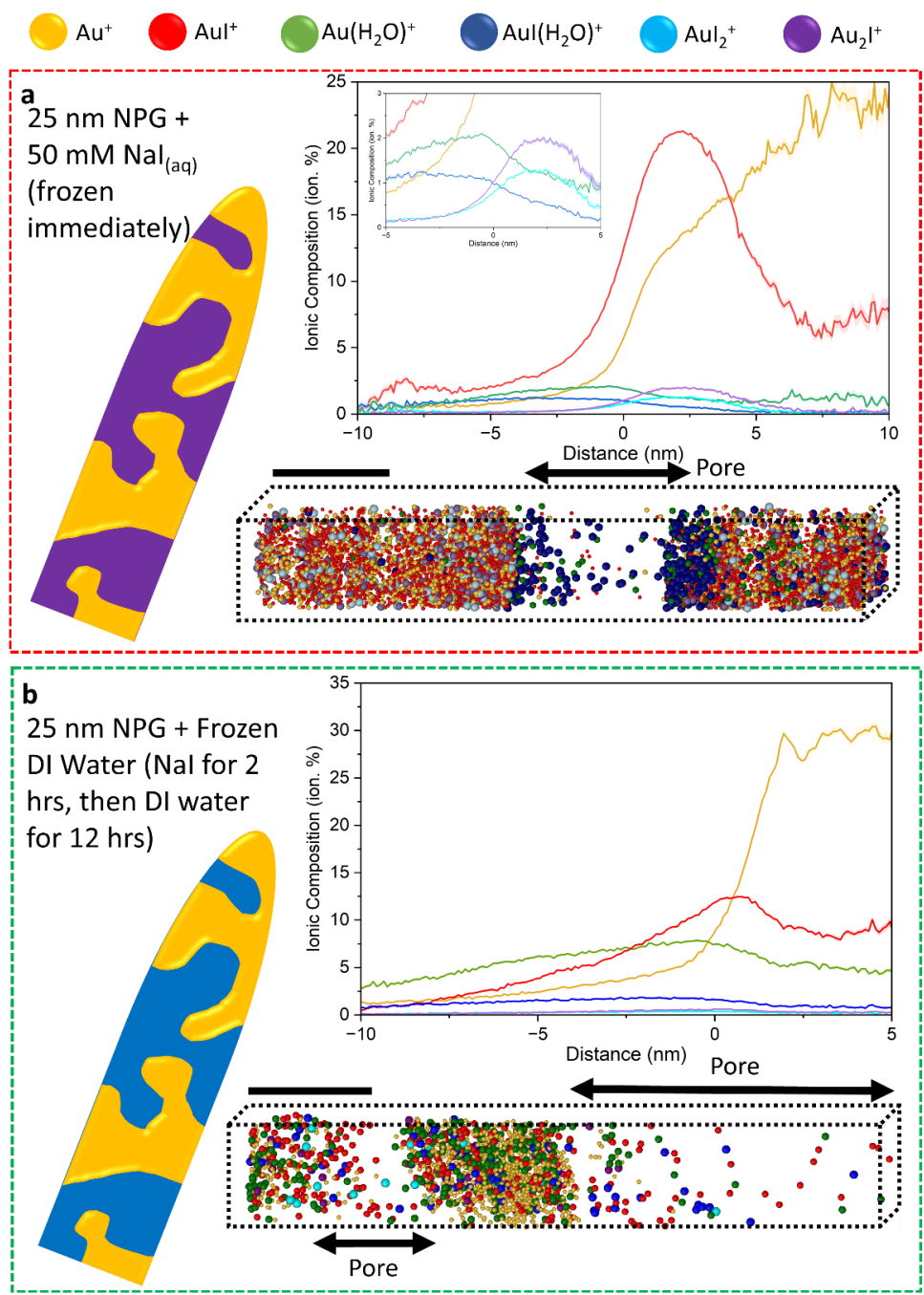

**Figure 5:a**) Proximity Histogram (proxigram) based on a 5% at Au isosurface, for the NPG sample, which was coarsened at 200 °C for 30 mins, and cryo-APT analysis carried out using 50 mM NaI$_{aq}$ electrolyte. The displayed species are the major complexes between Au and I. 3D APT reconstruction of a 5 nm-diameter cylindrical profile through a ligament/pore in the same sample as the proxigram in a). The species displayed are the major complexes between Au and I. The "empty" region indicates the pore but does not display the water species. (scale bar = 5 nm). **b)** Proximity Histogram (proxigram) based on a 5% at Au isosurface, for the NPG sample, which was coarsened at 200 °C for 30 mins, and cryo-APT analysis carried out after the substrate had been immersed in NaI for 2 hours and submersed in DI water overnight. The displayed species are the major complexes between Au and I. 3D APT reconstruction of a 5 nm-diameter cylindrical profile through a ligament/pore in the same sample as in c). The displayed species are the major complexes between Au and I., with the "empty" region indicating the pore and showing no water species (scale bar = 5 nm).

To validate our gold dissolution observations using another characterisation method, ICP-MS measurements were performed on 50 mM NaIaq solutions after 2 gold samples of the same size had been immersed in the solutions for 1 and 2 hours, respectively. Gold was detected in both solutions, with the absolute amounts dissolved in the two samples being 0.11 μg and 0.16 μg, respectively. This further confirms that the NPG samples are being etched in the presence of iodide ions, with gold leaching at neutral pH.  When considering the conditions chosen in our study, the dissolution of Au would most probably be mediated by Au-I complexation, forming Au-I species seen throughout the analysis, which then react with remaining iodide in the pores, as seen in previous studies looking at iodide-mediated Au dissolution [13,64]. While gold iodide is not expected to be soluble in water, iodide ions in the solution could facilitate the dissolution of AuI from the ligaments and explain its presence in ICP measurements [73].

As discussed extensively above, rather than  I⁺ (as ions in atom probe are detected as positive ions) being detected exclusively within the porous regions of the samples, their proxigram concentrations closely match those of gold in all three samples, deviating slightly at the ligament surface and, unexpectedly, extending fully into the shell of the nanoligaments. This clearly suggests that iodide causes etching/dissolving of the gold surface before freezing, which has been seen with the etching of gold nanostars in the presence of iodide [13] and at iodide concentrations ~10,000x less than here (50 mM vs 1 μM). The suggested mechanism for the dissolution of gold due to iodide on Au-rich nanoconfined surfaces is:

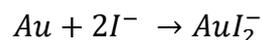

$$Au + 2I^- \rightarrow AuI_2^-$$

However, since we see significant amounts of Au₂I, it is likely that the reaction mechanism:

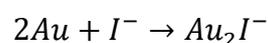

$$2Au + I^- \rightarrow Au_2I^-$$

And

$$Au + I^- \rightarrow AuI^-$$

The nature of the complexation may be greater than what appears here due to detection losses, but because the overall background signal in our samples was low (5 ppm/sec and 8 ppm/sec in **Figures 4a** and **4b,** respectively), we do not expect a substantial loss in $Au_xI_y$ ion signals. Ultimately, the ratio of liquid-solid-phase AuI complexes will depend on the iodide concentration available within the nanopores, as well as the concentrations of Au in the nanoporous metal and its precursor. Altering the concentration of iodide and Au shall affect the final thickness and chemistry of functionalized Au-I nanoligament shells.

## 3. Conclusion

In this study, we have revealed the interaction of iodide ions with nanoporous gold substrates using cryogenic atom probe tomography, enabling novel analysis of this system with high spatial and chemical resolutions.  Our results show the formation of AuI complexes in the presence of 50 mM sodium iodide, highlighting the affinity of Iodide ions on high-curvature gold surfaces. Atom probe tomography has allowed us to visualise the distribution of different ionic species within and near the nanoporous ligands, confirming their preferential attraction to gold. We also observed significant gold dissolution in the presence of iodide, as confirmed by ICP measurements, and the protrusion of iodide species into the gold ligaments, as captured by chemical mapping. Finally, despite the hugely challenging nature of cryo-APT analysis, we show that complex processes during chemical reactions can be studied using this technique.

These results provide novel insights into iodide-gold interactions and highlight the development of cryogenic-APT to analyse complex liquid-solid interfaces, and its application to image processes and systems that other techniques cannot observe due to limitations in spatial and chemical resolution. We also demonstrate progress in using cryo-APT to investigate electric double-layer science, but more work is needed before this can be reliably resolved.

## 4. Methods

## 4.1. Nanoporous Gold Fabrication

An $Ag_{80}Au_{20}$ at% alloy was annealed at 900 °C in an inert Ar atmosphere. Rectangular sections (10 mm × 3 mm) of the annealed alloy were cut and mechanically polished using 600 and 1200-grit SiC paper. The nanoporous gold substrates were prepared via chemical dealloying in 5 mL 68% $HNO_3$ (VWR Chemicals, TECHNICAL). After dealloying, the initial pore size was determined by SEM to be 15 nm (±8 nm). The substrates were then thermally annealed at 200 °C for 30 minutes to induce coarsening of the NPG ligaments, resulting in a pore size of 35 nm (± 15 nm). After annealing, the NPG was kept in 100 mL of 18 MΩ cm-1 de-ionised ultra-pure water overnight (Millipore Direct Q3) to allow residual dissolved silver to leach from the pores. For subsequent cryo-APT experiments, the samples were attached to an APT copper clip and connected to a low-profile Cameca puck suitable for UHV cryogenic transfer. It was quickly immersed in a 50 mM solution of NaI (Sigma-Aldrich, ACS reagent, ≥99.5), prepared in 18 MΩ cm−1 de-ionised water (Millipore Direct Q3) to keep the sample hydrated.

## 4.2. Sample Freezing, Transfer, and Cryogenic Specimen Fabrication

Ultra-high-vacuum cryogenic sample transfer is required to keep the sample frozen in its native state and to prevent frosting from residual water vapour. This transfer mechanism is detailed in previous cryo-APT studies, pioneered in the Laplace project at the Max-Planck-Institute for Sustainable Materials, in Dusseldorf [30]. A transfer suitcase maintained at ultra-high vacuum ($10^{-11}$

mBar) and cryogenic temperatures (≈ 90 K) (Ferrovac GMBH) was used to transport the frozen samples between an inert nitrogen glovebox, a Thermo Fisher Hydra/Helios dual beam PFIB (Xe plasma source) fitted with an Aquilos cryo-stage, and Cameca LEAP 5000 XR (Cameca Instruments).

The dealloyed samples immersed in the NaI solution were rapidly frozen inside the glovebox by plunging into liquid nitrogen. When excess liquid was present on the nanoporous substrate surface, blotting with lint-free tissue was performed before freezing. The frozen samples were transferred from the glovebox into the transfer suitcase via a load lock, which reached a vacuum of $10^{-6}$ mBar.

The "satellite dish" milling procedure, developed initially by Halpin [74] and subsequently used by El-Zoka et al.[33] was also used during this study to eliminate the need for a cryogenic lift-out procedure. Minor modifications introduced to the milling steps helped to reduce the time required to make each APT needle. Ion beam currents ranged from 1.3 μA to 0.1 nA with an accelerating voltage of 30 kV. No images were taken after this to avoid any sublimation of the ice induced by the electron beam. Fabrication time is between 90 to 120 minutes per tip, allowing for multiple tips to be fabricated in each session. By fashioning the tips this way, we create tips spanning microns in length that can be run for extended periods before fracturing.

## 4.3. Cryo-APT

APT analysis was carried out in a Cameca LEAP 5000 XR (Cameca Instruments) using Laser pulsing mode with a pulse of 60 pJ and pulse rates between 25-100 kHz. The detection rate ranged from 0.1% to 1% (0.001-0.01 ions per pulse). The stage temperature was kept constant at 50 K for all experiments. Reconstructions were carried out in AP Suite 6.1, IVAS 6.1.3.42.

Reconstructions were created using Ag as the primary ion at a field of 24 V/nm. Initial tip radii were dependent on each sample, based on the tip obtained during FIB milling.

We observe peaks corresponding to Cu and $CuOH_2$, which are impurities in the precursor alloy, confirmed by APT analysis (**Figure S10**) of the $Ag_{80}Au_{20}$ at% alloy. The impurity accounted for 0.035% of the total counts and had no meaningful impact on our data, as Cu ions would not affect dealloying at this concentration. It also revealed no clustering of Au and Ag atoms in the precursor alloy, indicating a homogeneous solid solution, and that the alloy's annealing was successful. The $CO_2^+$ peak arises from trapped $CO_2$ in the frozen water and small amounts of residual gas in the chamber.

Trajectory aberrations are expected to arise from mismatches in the evaporation field between the nanoligaments, the Au-I phase, and the frozen solution in the pores. This will lead to trajectory overlaps, affecting compositional estimates. This might explain the relatively low composition of frozen water around the ligaments, as atoms originating from the ligaments are subject to strong aberrations and are imaged as part of the pores, typical of the aberrations caused by differences in the electric field necessary for field evaporation[75]. This typical issue of trajectory overlaps will also cause our high-evaporation field nanoligaments to lead to an underestimation of the local composition[76], and an overestimation of the size. Despite these complexities, results demonstrate that our approach enables consistent analysis of solute ions and the products of Au-I chemical reaction.

## 4.4.   Other Methods (ICP, Non-cryo APT, EIS, CV)

**OCP**

OCP measurements were carried out using NPG samples as the working electrode, dealloyed as before, and annealed at 200 °C for 30 minutes. The same 50 mM $NaI_{aq}$ solution was used, with an Ag/AgCl (sat. KCl) reference electrode (name/make) and a Pt wire as the counter electrode on a Metrohm Autolab PGSTAT 302 potentiostat.

**ICP-MS**

ICP-MS measurements were conducted using an Agilent 8900 ICP-MS Triple Quadrupole, focusing on gold concentrations only. Solutions were made up of nanoporous gold substrates annealed at 200 °C for 30 mins and immersed in our 50 mM $NaI_{aq}$ solution for 30 mins and 120 mins.  Standards at concentrations of 0, 2, 10, 25, 50, and 100 ppb were prepared using an Au standard for ICP (Sigma-Aldrich Trace**CERT**®, 1 g/L Au in hydrochloric acid).

**APT**

APT was carried out on NPG samples annealed for 30 mins at 200 °C (~35 nm pore size), which were first immersed in 50 mM $NaI_{aq}$ for 120 mins, stored in ultrapure water overnight, and filled with copper via electrodeposition, similar to previous studies by El-Zoka et. al [65,77].  After the electrodeposition, APT needles were fashioned using a well-established lift-out procedure (**Figure S7**). APT was carried out in laser mode with a pulsing energy of 60 pJ, a pulse rate of 100 kHz and a detection rate of 0.5%. The stage temperature was kept at 50 K. Reconstructions and data analysis were carried out in AP suite 6.1, IVAS 6.1.3.42.

# 5. Acknowledgments

We thank Guangmeimei Yang for supporting our ICP experiments to measure gold dissolution. O.R.W acknowledges the support of the Centre for Doctoral Training in the Advanced Characterisation of Materials PhD programme and the BASF Coating division for funding of his PhD. A.A.Z acknowledges financial support from the Department of Materials at Imperial College London. The authors are thankful for access to the Imperial

Cryogenic Microscopy Centre at Imperial College London, which is supported by EPSRC grant no. **ep/v007661/1**.

## 6. Competing Interests

The authors declare there are no competing interests.

## Atomic-scale Imaging of Iodide-Gold Interactions in Nanoconfined Liquid-Solid Interfaces

Oliver R. Waszkiewicz[a], Yuxiang Zhou[a], Baptiste Gault [a,b,c], Finn Giuliani[a], Mary P. Ryan[*a], Ayman A. El-Zoka[*a, c]

[a] Department of Materials, Royal School of Mines, Imperial College London, Exhibition Road, SW7 2AZ, UK

[b] Max Planck Institute for Sustainable Materials, Max-Planck-Str. 1, 40239 Düsseldorf, Germany

[c] present address Univ. Rouen Normandie, INSA Rouen Normandie, CNRS, Groupe de Physique des Matériaux, UMR 6634, F-76000, Rouen, France

[*]Corresponding Author

**This file includes:**

Figs. S1 to S7 and references

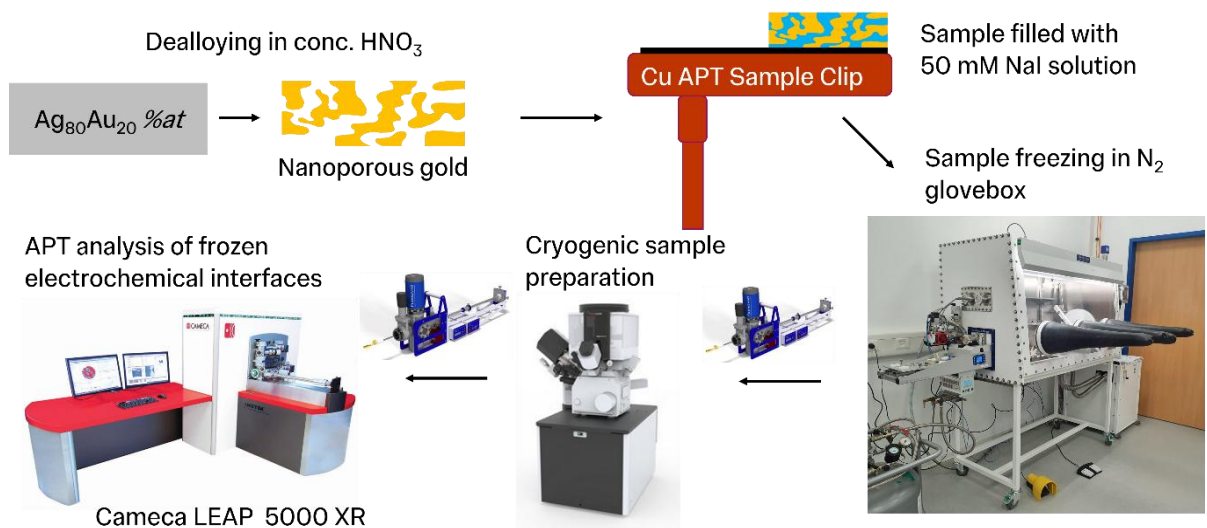

**Figure S1**: A workflow process for creating and transferring the NPG samples under cryogenic, ultra-high vacuum conditions to conduct cryo-APT analysis[1].

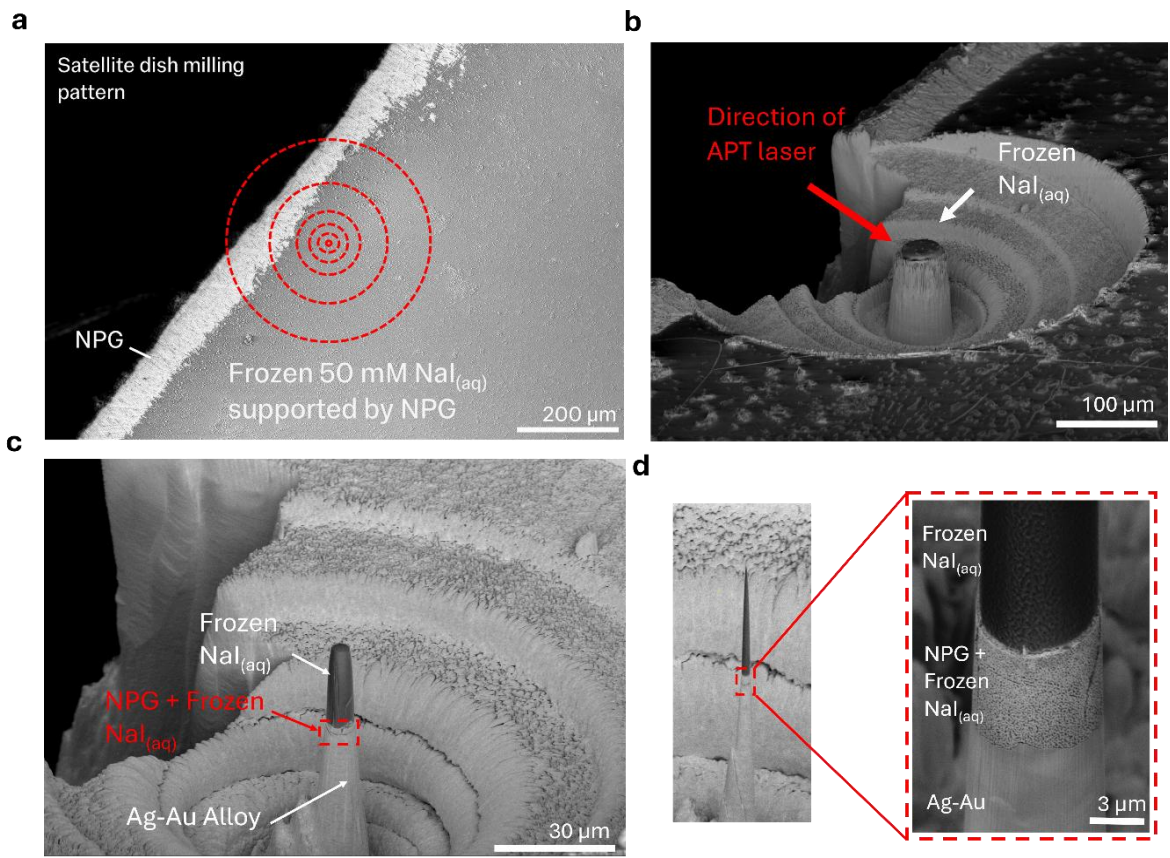

**Figure S2**: Focused ion beam (FIB) milling of frozen aqueous NaI supported by nanoporous gold (NPG) for atom probe tomography (APT). **A)** Low-magnification SEM image showing the satellite dish milling pattern (red dashed circles) used to approach the analysis region containing frozen 50 mM NaI(aq) supported by NPG, taken after a few milling steps. **B)** Intermediate magnification revealing the frozen NaI(aq) pillar after coarse milling, with the APT laser incidence direction indicated (red arrow), close to the end stage of milling. **C)** Higher magnification view showing the layered structure of frozen NaI(aq), NPG + frozen NaI(aq), and underlying Ag–Au alloy. **D)** *(left)* Image showing the tip near the final geometry of the APT needle. *(right)* High magnification of the tip apex showing the distinct regions of frozen NaI(aq), NPG infiltrated with frozen NaI(aq), and the Ag–Au base [2,3].

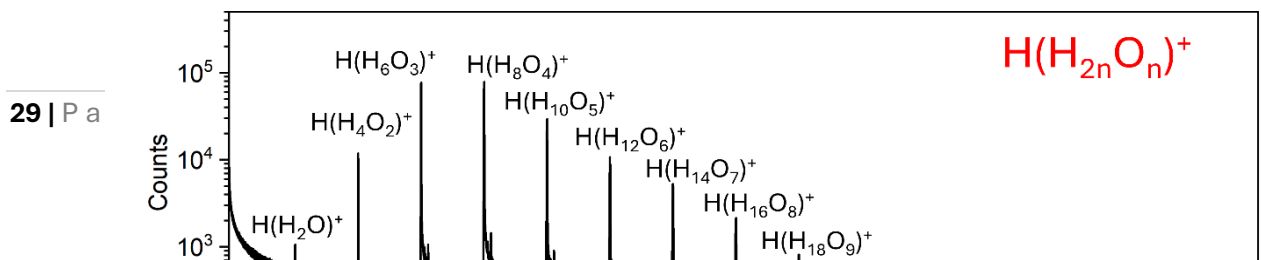

**Figure S3**:  Mass spectrum showing mass-to-charge state ratio from 0-300 Da, on a logarithmic plot, collected in the region of bulk electrolyte on top of the NPG sample from **S2**. The major peaks and peaks of interest have been labelled with their respective ions.

The formula for the successive water cluster identity is:

$$Water\ Cluster = H(H_{2n}O_n)$$

Starting from N=1 ($H(H_2O)^+$), excluding $H_2O$ (only appearing withing the NPG regions)[3,4].

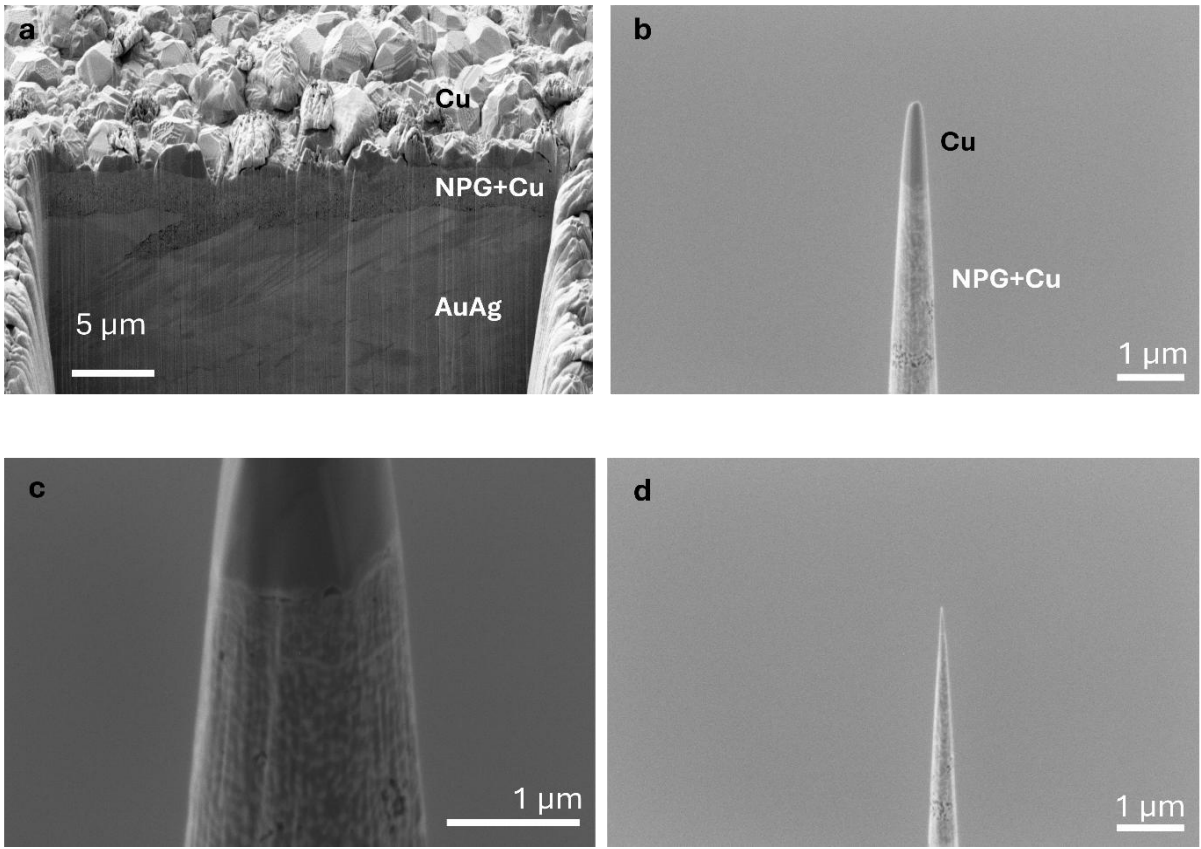

**Figure S4**: **a**) FIB cross-section of the 25 nm NPG sample filled with electrodeposited Cu after being submersed in 50 mM NaI for 2 hours. **B,C,D**) APT needle sharpening of the same sample from **A**). The lighter colour indicates NPG, and the darker colour Cu due to the higher atomic weight of Au for the backscattered electrons

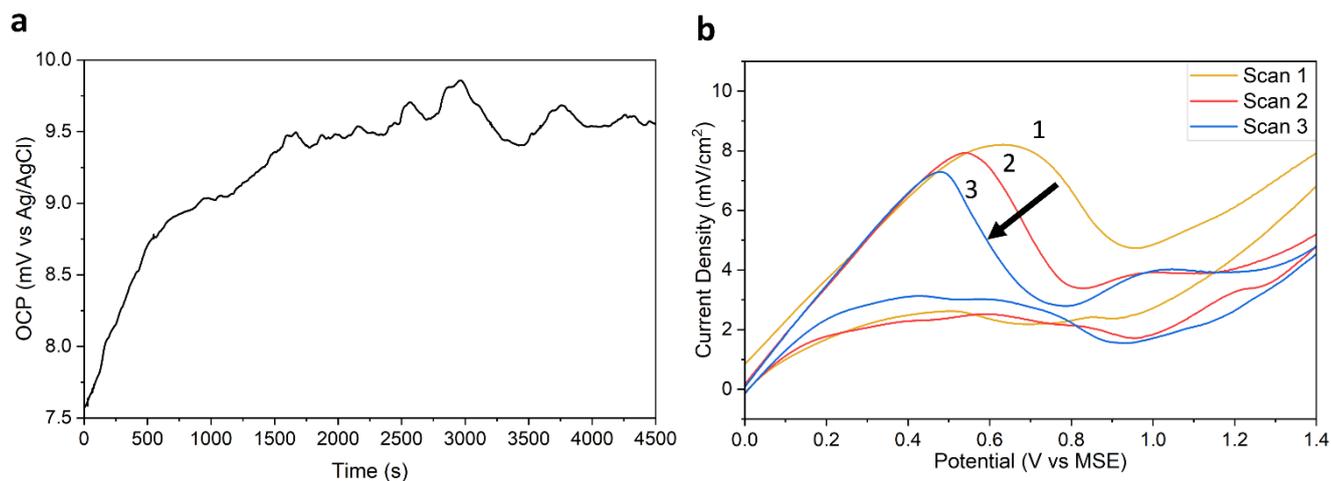

**Figure S5**: **a**) Open circuit potential measurement vs an Ag/AgCl reference electrode, a platinum counter electrode and the 25 nm NPG as a working electrode in a 50 mM NaI$_{(aq)}$ electrolyte. **b)** Cyclic voltammogram using the same setup as in **a**), with a scan rate of 50 mV/s, with the selected region shown being 0-1.4 V vs Ag/AgCl.

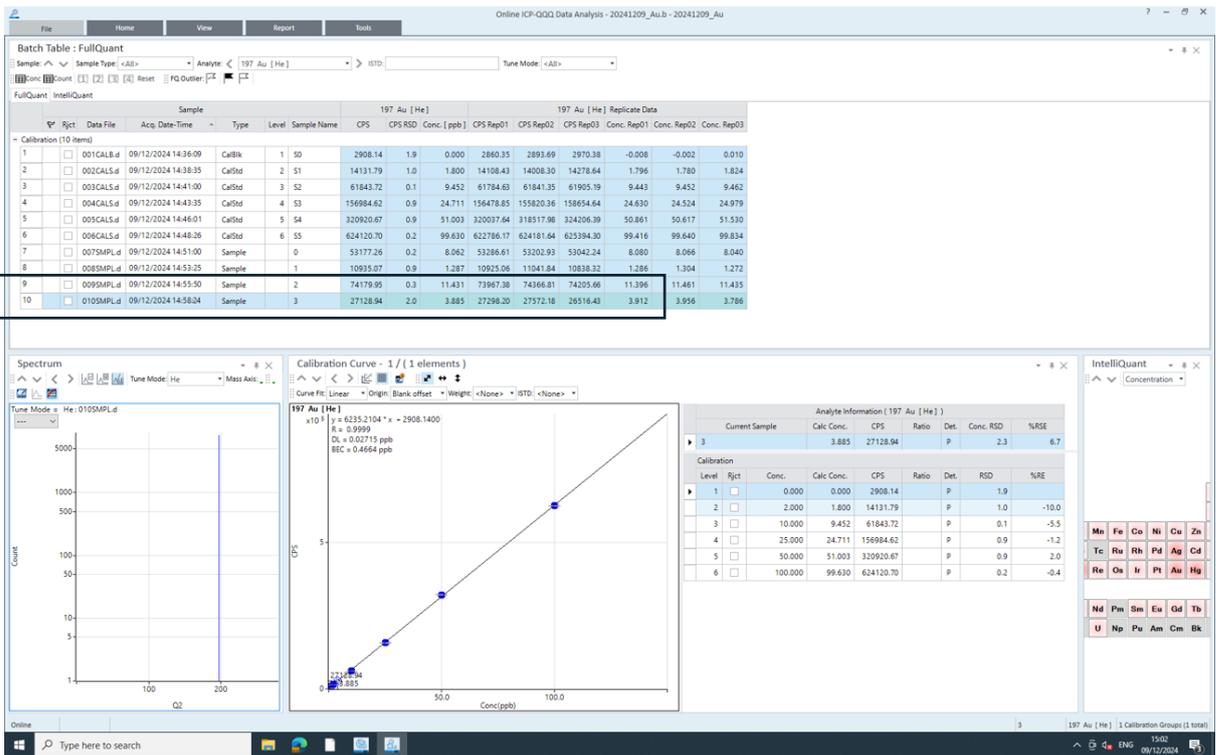

**Figure S6**: ICP analysis showing calibration curve, sample measurement and quantitative analysis for detection of Au, from dissolution of Au from 25 nm NPG, after it was submersed in 50 mM NaI for 60 min and 120 min.

 Samples 2 and 3 in the black box denote the concentration of Au detected after 60 min and 120 min of submersion in NaI, respectively. The absolute value of Au detected was 0.11 and 0.16 μg, as the total volume of solutions for 60 min and 120 min was different (10 mL vs. 50 mL, respectively).

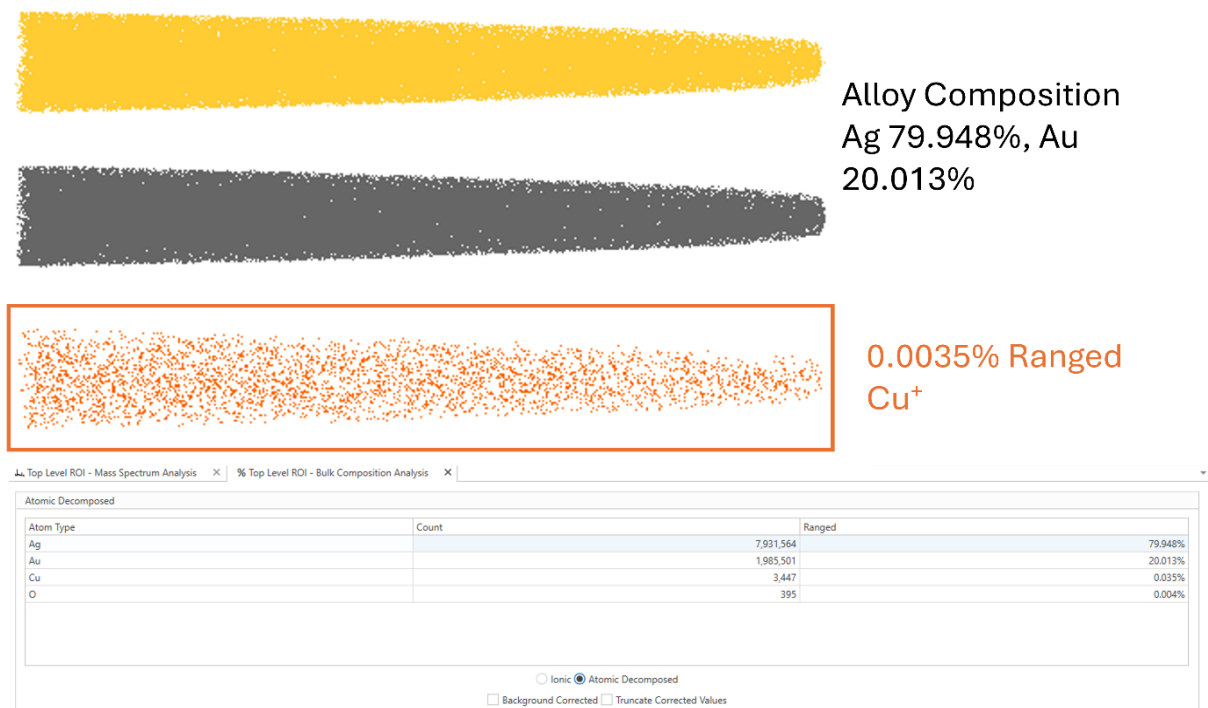

**Figure S7**: APT Reconstructions and ranged composition from the precursor AgAu alloy.